\documentclass[twocolumn]{IEEEtran}
\usepackage{cite}
\usepackage{wrapfig,graphicx,booktabs,fancyhdr,amsfonts}
\usepackage{bm,amssymb,amsmath,amsthm,wasysym,multirow,dsfont}
\usepackage{subcaption,algorithm,xcolor,pifont,float,enumitem}
\usepackage[noend]{algorithmic}

\newcommand{\vb}{\boldsymbol}

\newcommand{\ba}{\begin{array}}
\newcommand{\ea}{\end{array}}

\DeclareMathAlphabet{\mathpzc}{OT1}{pzc}{m}{it}

\begin{document}
\title{Maximum-Likelihood Power-Distortion Monitoring for GNSS Signal Authentication}
\author{Jason~N.~Gross,~\IEEEmembership{Member,~IEEE,}~Cagri~Kilic,~and~Todd~E.~Humphreys,~\IEEEmembership{Member,~IEEE}}

\maketitle

%\fancyhf{} % sets both header and footer to nothing
%\renewcommand{\headrulewidth}{0pt} \pagestyle{plain}
%\thispagestyle{fancy} % Requires the package fancyhdr
%\rfoot{\footnotesize \bf Preprint of the 2013 IEEE Global Conference on Signal
%  and \\Information Processing (GlobalSIP), Austin, TX, Dec 3--5, 2013}
%\lfoot{\footnotesize \bf Copyright \copyright~2013 by Kyle D. Wesson, Jason N. Gross, Brian
%  L. Evans, \\ and Todd E. Humphreys}

\begin{abstract}
  We propose an extension to the so-called PD detector. The PD detector jointly monitors received power
  and correlation profile distortion to detect the presence of GNSS carry-off-type spoofing,
  jamming, or multipath. We show that classification performance can be
  significantly improved by replacing the PD detector's
  symmetric-difference-based distortion measurement with one based on the
  post-fit residuals of the maximum-likelihood estimate of a single-signal
  correlation function model. We call the improved technique the PD-ML
  detector. In direct comparison with the PD detector, the PD-ML detector
  exhibits improved classification accuracy when tested against an extensive
  library of recorded field data. In particular, it is (1) significantly more
  accurate at distinguishing a spoofing attack from a jamming attack, (2)
  better at distinguishing multipath-afflicted data from interference-free
  data, and (3) less likely to issue a false alarm by classifying multipath as
  spoofing. The PD-ML detector achieves this improved performance at the
  expense of additional computational complexity.
\end{abstract}

\begin{IEEEkeywords}
  satellite navigation systems, Global Positioning System, navigation
  security, GNSS spoofing, GNSS jamming, GNSS authentication
\end{IEEEkeywords}

\section{Introduction}
\IEEEPARstart{C}{ivil} GNSS receivers are susceptible to counterfeit spoofing
signals~\cite{t_humphreys_gcs08,shepard2012_IJCIP,
  kerns2014unmanned,bhatti2017hostile}. To defend against spoofing attacks,
several algorithms have been developed and reported in the literature. A
comprehensive review of GNSS spoofing detection methods is provided
in~\cite{psiaki2016gnssSpoofing}, where methods are broadly categorized as (1)
cryptographic techniques, (2) geometric techniques that exploit
angle-of-arrival diversity, or (3) signal processing techniques that do not
fall into categories (1) or (2), such as the PD detector~\cite{wesson2018pincer}.

Within the latter category, the subset of techniques which do not require
additional hardware and can be implemented via a firmware update are
particularly attractive for widespread adoption.  Of these, the PD detector, which jointly
monitors received power and correlation profile distortion, 
has been shown to reliably alarm in the presence of carry-off-type spoofing or jamming
attacks while maintaining a low multi-channel false alarm rate, when tested against 27 high quality data recordings~\cite{wesson2018pincer}.

An appealing attribute of the PD detector is simplicity: its received power
and correlation profile distortion measurements are computationally
lightweight and amenable to analysis.  However, when tested
in~\cite{wesson2018pincer} against an extensive set of empirical data, the PD
detector was shown to incorrectly classify most instances of spoofing as
jamming, and its $0.57$\% single-channel false alarm rate (false declaration
of interference-free or multipath-afflicted signals as spoofing or jamming),
is too high for some applications of practical interest. This correspondence article extends
the PD detector to address these shortcomings, and for compactness, it is intended to be read alongside
the original PD detector paper~\cite{wesson2018pincer}. 

The PD detector's shortcomings stem from two undesirable features of the
symmetric difference (SD) as a measure of distortion.  First, the standard SD
involves only a pair of complex correlation function taps and is not
particularly sensitive to correlation function distortion not aligned with
these taps.  Thus, if the SD taps are closely spaced around the prompt
correlation tap, then the SD can be a good distortion measure during the
initial stage of a carry-off-type spoofing attack, but quickly becomes
insensitive as the attack proceeds.  Second, the SD depends on the receiver's
code tracking loop to align its pair of taps symmetrically about the authentic
correlation peak.  But even in the absence of spoofing or significant
multipath, thermal noise prevents the code tracking loop from achieving
perfectly symmetric alignment.  As a consequence, the SD tends to exaggerate
the actual distortion under the null (thermal noise only) and jamming
hypotheses.

The PD-ML detector's distortion measure avoids both of these limitations.  It
begins by measuring the correlation function at many more than two taps, which
offers two advantages: (1) the PD-ML detector can be made sensitive to
distortion far from the prompt tap, thus extending the range of
spoofing-to-authentic code offsets over which it can detect spoofing, and (2)
the PD-ML detector's measure of distortion can be made independent of the
receiver's code and carrier tracking loops in the sense that small tracking
errors in no way affect the reported distortion. This is done by exploiting
data from the additional taps to obtain the maximum-likelihood estimates of
the amplitude, code phase, and carrier phase of the correlation peak in a
single-signal correlation function model. After removing this estimated
correlation function model from each of the correlation taps, the PD-ML
detector takes as its distortion measure the squared magnitude of the
normalized post-fit residuals. As will be shown, distortion measured in this
way permits more accurate classification of jamming, multipath, and spoofing
than distortion based on a simple symmetric difference.

Herein, we adopt, without alteration, several aspects of the PD detector~\cite{wesson2018pincer}. 
In particular, the signal models (Section II of~\cite{wesson2018pincer}), probability distributions (Section III of~\cite{wesson2018pincer}), and the
Monte-Carlo-type Bayes-optimal decision rule design strategy developed (Section VI of~\cite{wesson2018pincer}). This correspondence, a significant extension of our
conference paper~\cite{gross2017gnss}, exploits multi-tap maximum-likelihood
multipath estimation and demonstrates the proposed PD-ML detector's advantages
through a direct comparison with the original PD detector on the same
experimental data recordings and using the same Bayes-optimal detector design
strategy as presented in~\cite{wesson2018pincer}. Details of the PD-ML detector
are presented in Section \ref{sec:meas-models}, followed by simulated (Section
\ref{sec-rule}) and experimental (Section \ref{sec-results}) evaluation of its
classification performance.  All source code required to generate decision
rules using the PD-ML detector is publicly available at
https://github.com/navSecurity/P-D-defense.

\section{Signal Models}
\label{sec:sig-models}
For a thorough discussion of the GNSS signal models adopted in this work,
starting with the pre--correlation signal model, the reader is directed to the
paper on the PD detector~\cite{wesson2018pincer}. However, to provide the
context necessary to understand this paper's proposed extension to the PD
detector, a brief review of the single-interferer post-correlation model
assumed in~\cite{wesson2018pincer}, which is adapted from \cite{vannee1993ssme},
is presented here. This model gives the complex-valued receiver correlation
function at some arbitrary code offset, or lag, $\tau$, as
\begin{align}
  \label{eq:af}
  \xi_k(\tau) = \beta_k [ \xi_{{\rm A}k}(\tau) + \xi_{{\rm I}k}(\tau) + \xi_{{\rm N}k}(\tau) ]
\end{align}
where $\beta_k$ is the average value of the automatic gain control scaling
factor over the $k$th accumulation interval, and $\xi_{{\rm A}k}(\tau)$,
$\xi_{{\rm I}k}(\tau)$, $\xi_{{\rm N}k}(\tau)$ are the complex correlation
function components corresponding to the authentic signal, the interference
signal, and thermal noise, respectively.

Just as in~\cite{wesson2018pincer}, the correlation components for the authentic
and interference signal, $\xi_{{\rm A}k}(\tau)$ and $\xi_{{\rm I}k}(\tau)$,
are assumed to be modeled as
\begin{align*}
  \xi_{{\rm A}k}(\tau) & = \sqrt{P_{{\rm A}k}} R(-\Delta \tau_{{\rm A}k} + \tau) \exp(j
  \Delta \theta_{{\rm A}k}) \\
  \xi_{{\rm I}k}(\tau) & = \sqrt{\eta_kP_{{\rm A}k}}
  R(-\Delta \tau_{{\rm I}k} + \tau) \exp(j \Delta \theta_{{\rm I}k})
\end{align*}
where $R(\tau)$ is the GNSS auto-correlation function, and, at the $k$th
accumulation interval, $P_{{\rm A}k}$ is the average value of the authentic
signal's power, $\eta_k$ is the average interference signal's power advantage
over the authentic signal (i.e., $\eta_k=P_{{\rm I}k}/P_{{\rm A}k}$), and
$\Delta \tau_{{\rm A}k}$ is the average value of the code offset
$\tau_{\rm A} - \hat{\tau}$, with $\tau_{\rm A}$ being the true code phase of
the authentic signal and $\hat{\tau}$ being the receiver's estimate of the
same. Similar definitions follow for $\Delta \tau_{{\rm I}k}$,
$\Delta \theta_{{\rm A}k}$, and
$\Delta \theta_{{\rm I}k}$~\cite{wesson2018pincer}.

The thermal noise component of the correlation function,
$\xi_{{\rm N}k}(\tau)$, is modeled as having independent in-phase (real) and
quadrature (imaginary) components, each modeled as a zero-mean Gaussian white
discrete-time process:
\begin{align*}
  E[\mathds{R}\{\xi_{{\rm N}k}(\rho)\} \mathds{I} \{\xi_{{\rm N}j}(\nu)\}] = 0 \quad \forall
  ~\rho,\nu,k \neq j
\end{align*}
As discussed in~\cite{dierendonck1992ncs}, thermal noise is correlated in the
lag domain.  For samples within $2\tau_c$ of each other, where $\tau_c$ is the
chip-width of the GNSS signal, correlation in $\xi_{{\rm N}k}(\tau)$ is
modeled as
\begin{align*}
  E[\xi_{{\rm N}k}(\rho) \xi^*_{{\rm N}k}(\nu)] = \left\{\ba{ll} 2\sigma_{\rm N}^2(1-
  |\rho-\nu|/\tau_c) & |\rho-\nu|\leq 2\tau_c\\
   0 &  |\rho-\nu| > 2\tau_c \ea \right.
\end{align*}
Here, $^*$ denotes the complex conjugate and $\sigma_{\rm N}^2$ is the
variance of the in-phase and quadrature components of $\xi_{{\rm N}k}(\tau)$,
which is related to the spectral density of a white noise process that is
modeled as the sum of two independent components associated with thermal noise
and multi-access noise.  The spectral density of the thermal noise, $N_0$, is
assumed to be constant whereas the spectral density of the multi-access noise,
$M_0$, is assumed to be variable, as detailed in~\cite{wesson2018pincer}. When
averaging over an accumulation interval $T$ the variance is given as
$\sigma_{\rm N}^2 = (N_0 + M_0)/2T$.

\section{Measurement Models}
Like the PD detector, the PD-ML detector simultaneously monitors received
power and correlation function distortion.  Its received power monitor is
identical to the PD detector's, as described in ~\cite{wesson2018pincer}, but
its distortion monitor differs considerably.  This section develops the
measurement model for the PD-ML detector's distortion monitor.
 
The PD-ML detector models the in-phase and quadrature (IQ) samples of the
correlation function $\xi_k$ as a function of only the authentic signal and
thermal noise, neglecting $\xi_{{\rm I}k}$ in (\ref{eq:af}).  It employs a
maximum-likelihood estimator to estimate the authentic signal's amplitude,
code phase, and carrier phase, and takes the squared magnitude of the
normalized post-fit residuals as its measure of correlation profile
distortion.

Maximum-likelihood estimation based on GNSS correlation data is
well-established in the literature~\cite{van1994multipath, townsend1995performance, 
lentmaier2006maximum, sahmoudi2008fast,blanco2012multipath}.  It is routinely employed within high-end
GNSS receivers for multipath mitigation. This paper's distortion monitor
adapts the particular maximum-likelihood estimator developed in
~\cite{blanco2012multipath}, as described subsequently, though other approaches could be adopted in the PD-ML detector in a straight-forward manner.

\label{sec:meas-models}
\subsection{Multi--Tap Maximum--Likelihood Correlation Function Estimator}
Let $l$ be the number of signal taps devoted to maximum-likelihood estimation.
For convenience, we assume that $l$ is odd and that taps are distributed, so the centermost
tap is approximately aligned with the receiver's estimated correlation
function peak and the remaining taps are spaced evenly across the range
$\pm\tau_c$. The uniform tap interval is
\begin{align*}
 \Delta\delta = \frac{2\tau_c}{(l-1)}
 \end{align*} 
 and the $l\times 1$ vector of tap locations is given by
 \begin{align*}
   \vb{\delta}=\begin{bmatrix}-\tau_c, -\tau_c+\Delta\delta, \hdots, \tau_c-\Delta\delta, \tau_c\end{bmatrix}^T
 \end{align*}
 with $\delta_i = -\tau_c + (i-1) \Delta \delta$ representing the $i$th tap
 location, $i = 1, \dots, l$.
 
 Modeling the correlation function $\xi_k(\tau)$ as interference-free (i.e.,
 ${\xi_{{\rm I}k}}(\tau)=0$ for all $\tau$), the complex-valued $i$th tap at
 time index $k$ is expressed in terms of the authentic signal's
 gain-controlled amplitude ${a}_{{\rm A}k}$, carrier phase
 ${\phi}_{{\rm A}k}$, and code phase ${\tau}_{{\rm A}k}$ as
 \begin{align}
   \label{eq:singleSignalModel}
    {\xi}_k(\delta_i) & =\beta_k [ \xi_{{\rm A}k}(\delta_i) + \xi_{{\rm
                        N}k}(\delta_i) ] \nonumber \\ 
    & ={a}_{{\rm A}k}\exp({j{\phi}_{{\rm A}k})}R(\delta_i-{\tau}_{{\rm A}k})+ \beta_k\xi_{{\rm
        N}k}(\delta_i)
\end{align}
The parameters $\{{a}_{{\rm A}k}, {\tau}_{{\rm A}k}, {\phi}_{{\rm A}k}\}$ are
estimated according to this model by an adaptation of the maximum likelihood
technique in~\cite{blanco2012multipath}.  This approach separates estimation
of the code phase from estimation of amplitude and carrier phase by exploiting
the linear relationship
\begin{equation}
\vb{\xi}_k={H}({\tau}_{{\rm A}k}, \vb{\delta}) {a}_{{\rm A}k}\exp({j{\phi}_{{\rm A}k}})
\end{equation}
where
$\vb{\xi}_k = [ {\xi_k}(\delta_1),\dots,
{\xi_k}(\delta_l)]^T$ and the observation matrix
${H}({\tau}_{{\rm A}k}, \vb{\delta})$ is
\begin{equation}
  {H}({\tau}_{{\rm A}k},\vb{\delta})=
  \begin{bmatrix} R(\delta_1-{\tau}_{{\rm A}k}) \\
    \vdots  \\    
    R(\delta_l-{\tau}_{{\rm A}k})  \end{bmatrix}
\end{equation}
First, a coarse search is performed by setting the code phase estimate
$\hat{\tau}_{{\rm A}k} = \delta_i$ for $i=1,\dots,l$ and, for each candidate
$\hat{\tau}_{{\rm A}k}$, solving for the maximum-likelihood estimate of
${a}_{{\rm A}k}\exp({j{\phi}_{{\rm A}k}})$:
\begin{align} \hat{a}_{{\rm A}k}& \exp({j\hat{\phi}_{{\rm
                              A}k}}) =  \nonumber \\
  & \left[{H}^T(\hat{\tau}_{{\rm
      A}k},\vb{\delta}){Q}^{-1}{H}(\hat{\tau}_{{\rm
      A}k},\vb{\delta})\right]^{-1}{H}^T(\hat{\tau}_{{\rm
      A}k},\vb{\delta}){Q}^{-1}\vb{\xi}_k
\label{LS}
\end{align}
where ${Q}$ is the $l\times l$ Toeplitz matrix that accounts for the
correlation of the complex Gaussian thermal noise among the
taps~\cite{blanco2012multipath}. The $(a,b)$th element of ${Q}$ is
${Q}_{a,b}=R(|a-b|\Delta\delta)$, where $\Delta\delta$ is the tap spacing.

The cost $J_k$ corresponding to each set of estimates
$\{\hat{a}_{{\rm A}k}, \hat{\tau}_{{\rm A}k}, \hat{\phi}_{{\rm A}k}\}$, is
calculated as
\begin{equation}
  J_k=\left\|\vb{\xi}_k-{H}(\hat{\tau}_{{\rm A}k},\vb{\delta})\hat{a}_{{\rm A}k}\exp({j\hat{\phi}_{{\rm A}k}})\right\|^2_{{Q}}
 \label{leastSquares}
\end{equation}
where the norm is defined such that
$\|\vb{x} \|_Q^2 = \vb{x}^T Q^{-1} \vb{x}$.  The two sets of estimates
yielding the smallest cost $J_k$ are extracted.  Because the cost $J_k$ is
proportional to the negative log likelihood function, the set with the minimum
cost is the maximum likelihood estimate.

\begin{figure}[ht]
 	\centering
\includegraphics[width=0.5\textwidth]{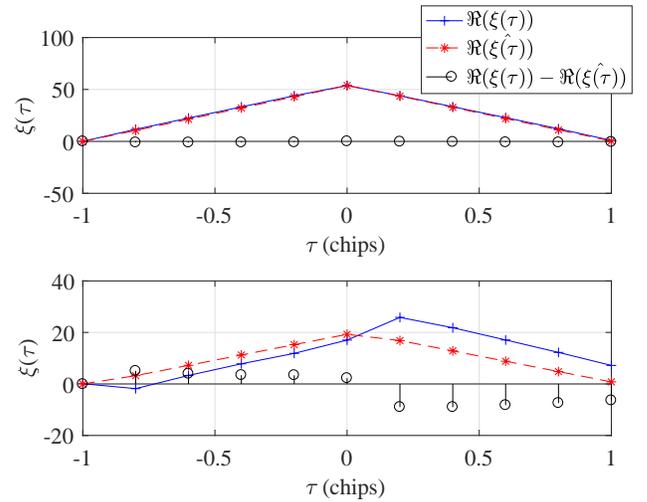}
\caption{Multi-tap samples of the in-phase components of the correlation function 
$\vb{\xi}(\tau)$, the maximum-likelihood estimation of $\hat{\vb{\xi}(\tau)}$ at the tap locations, 
and the difference between them for a nominal scenario (top panel) and a spoofing scenario (bottom panel).
For readability, only the in-phase components are shown.  The norm of residuals between the measured and 
estimated correlation function defines $J_k$, and is much larger when the single-signal assumption is violated (e.g., spoofing or multipath).}
\label{fig:graphicRepresentation}
\end{figure}

In a second step, a refined code phase estimate is obtained by a bisecting
search between the two lowest-cost code phase estimates.  At each bisection
point, new amplitude and carrier phase estimates are determined by
re-evaluating (\ref{LS}).  The process is repeated until $J_k$ is no longer
significantly reduced.  Upon convergence, the resulting estimates are accepted
as the maximum-likelihood estimates
$\{\hat{a}_{{\rm A}k}, \hat{\tau}_{{\rm A}k}, \hat{\phi}_{{\rm A}k}\}$, and
the corresponding $J_k$ is taken as the distortion measurement $D_k$.  A small
$J_k$ indicates that the single-signal model (\ref{eq:singleSignalModel})
accurately fits the correlation function data at time $k$; a large $J_k$
indicates the opposite, suggesting that multipath or spoofing interference is
present.  This is illustrated in Fig.~\ref{fig:graphicRepresentation}, where the top panel
illustrated a nominal scenario and the bottom panel represents a spoofing scenario.

This distortion metric is more informative than the symmetric
difference metric used in \cite{wesson2018pincer} because it (1) exploits data
from $l$ taps, whereas the symmetric difference only uses two, and (2) is
insensitive to noise- or dynamics-induced misalignment of the prompt tap
(located at $\delta_i = 0$) with the authentic signal peak, whereas the
symmetric difference falsely reports distortion in this circumstance.

\section{Decision Rule}
\label{sec-rule}
Design of the PD-ML's decision rule follows the same procedure outlined in the
PD detector~\cite{wesson2018pincer}, but with the new distortion metric
replacing the symmetric difference. First, we simulate the post-correlation
model (\ref{eq:af}) using the same model priors, parameter probability
distributions, and probability transition mechanisms for each hypothesis as
in~\cite{wesson2018pincer}. The hypotheses are denoted $H_i,~ i\in \mathcal{I}$
for $\mathcal{I} = \{0,1,2,3\}$, where the null hypothesis $H_0$ corresponds
to the interference-free case, and $H_i,~i = 1,2,3$ correspond respectively to
multipath, spoofing, and jamming. For each Monte-Carlo sample under each
hypothesis, a power measurement is made as in ~\cite{wesson2018pincer} and a
distortion measurement $D_k = J_k$ is made as described above.

\begin{figure}[ht]
  \centering
\includegraphics[width=0.5\textwidth]{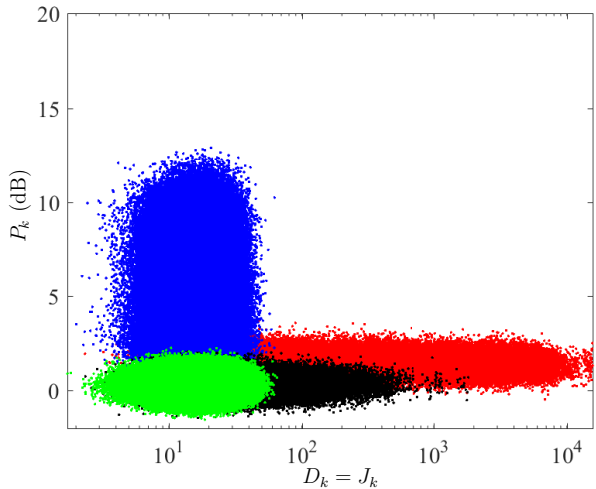}

\caption{Monte-Carlo-simulated distortion $D_k$ and received power $P_k$
  measurements for no interference ($H_0$, green), multipath ($H_1$, black),
  spoofing ($H_2$, red), and jamming ($H_3$, blue) based on a total of
  $N_{\rm P} = 10^5$ hypothesis samples and $N_{\rm M} = 20$ independent
  simulated measurements per sample, with $l = 11$ correlation taps
  contributing to the measurement of $D_k$.  As in ~\cite{wesson2018pincer},
  this simulation assumes a stealthy low-power-advantage-spoofer, which
  explains why the red points are clustered at low power relative to the blue
  (jamming) points.  Note that $D_k$ is plotted on a logarithmic scale.}
\label{MC11taps}
\end{figure}

\begin{figure}[ht]
 	\centering
\includegraphics[width=0.5\textwidth]{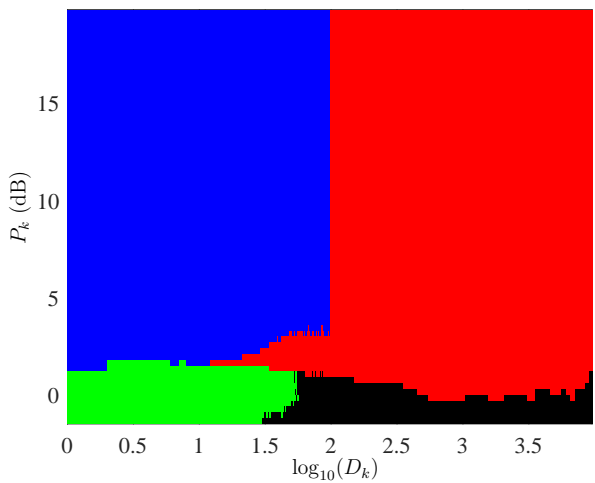}
\caption{Optimum decision regions for the $\theta$-dependent cost
  $C[i,\theta]$: $\Gamma_0$ (no interference, green), $\Gamma_1$ (multipath,
  black), $\Gamma_2$ (spoofing, red), and $\Gamma_3$ (jamming, blue).}
\label{fig:regions}
\end{figure}

Fig.~\ref{MC11taps}. shows the power-distortion measurements from a
Monte-Carlo simulation with $l = 11$ taps.  One can observe the following
behavior of the new distortion metric $D_k$:
\begin{itemize}
\item{ Under $H_0$ (interference-free): $D_k$ is small, as only thermal noise
    is present.}
\item{ Under $H_1$ (multipath-afflicted): $D_k$ is similar to that of the
    $H_0$ case when multipath is weak but attains a significant magnitude as
    multipath severity increases.}
\item{ Under $H_2$ (spoofing): $D_k$ overlaps with that of multipath ($H_1$),
    but exhibits a much wider range.}
\item{Under $H_3$ (jamming): $D_k$ is nearly indistinguishable from distortion
    under $H_0$, as jamming has little effect on the gain-controlled
    correlation function.}
 \end{itemize}
Based on the simulated data shown in Fig. \ref{MC11taps}, decision regions
corresponding to a Bayes-optimal decision rule for a parameter-dependent cost
$C[i,\theta]$ were found as described in ~\cite{wesson2018pincer}.
Fig.~\ref{fig:regions} shows the optimum decision regions $\Gamma_i$ for $i
\in \mathcal{I}$.  

Figs.~\ref{MC11taps} and~\ref{fig:regions} are analogous to Figs. 6 and 7 within~\cite{wesson2018pincer}.
When comparing these figures for PD and PD-ML, the most notable difference being that the PD-ML's distortion metric extends over a much larger domain than the PD  detector (i.e., 
notice the logarithmic scale in Figs. ~\ref{MC11taps} and ~\ref{fig:regions}). This increased distortion sensitivity could potentially lead to better discrimination between jamming and spoofing and, likewise, between interference free and multipath-afflicted.

Table \ref{tab:misclassSim1} shows classification statistics as evaluated by
applying the Bayes-optimal rule to a validation set of Monte-Carlo-generated
measurements that used the same assumptions as the distributions shown in Fig. \ref{MC11taps}.  This table reveals
that the PD-ML detector exhibits similar theoretical performance to the PD
detector.  Both tend to misclassify multipath ($H_1$) as interference free
($H_0$) because of the low cost assigned to this error when multipath is mild.
Otherwise, the PD-ML detector, like the PD detector, exhibits high detection
power and a low false alarm rate.

\begin{table}[h]
	\centering
	\caption{Simulation-evaluated classification for the decision regions
          in Fig. \ref{fig:regions}.  The table's $(i,j)$th element is the
          relative frequency with which the detector chose $i$ when $j$ was
          the true scenario.}
		\begin{tabular}[c]{ccccc}
		\toprule
		Decision & \multicolumn{4}{c}{True Scenario} \\
		\cmidrule(r){1-1} \cmidrule(r){2-5}
		& $H_0$ & $H_1$ & $H_2$ & $H_3$ \\  \midrule
		$H_0$ & 0.9947& 0.9083 & 0.0670 & 0.0039 \\
		$H_1$ & 0 & 0.0698 & 0.0117 & 0\\
		$H_2$ & 0.0043 & 0.0214 & 0.8463 & 0.0155 \\
		$H_3$ & 0.0010 & 0.0005 & 0.0750 & 0.9806 \\
		\bottomrule
	\end{tabular}	\label{tab:misclassSim1}
\end{table}

The sensitivity of the PD-ML detector's performance to the number of taps $l$
was explored.  Starting from $l = 41$ taps, the entire design process was
repeated for several choices of $l$ of decreasing value.  Performance
generally decreased with reduced $l$.  It was found that at least $l = 11$
taps were needed to maintain a level of theoretical detector performance
comparable with that of the PD detector (Table I of~\cite{wesson2018pincer}).

\section{Experimental Data}
\label{sec:experimental-data}
An independent evaluation of PD-ML was carried out against 27 empirical GNSS
data recordings, including 6 recordings of various spoofing scenarios~\cite{humphreys2012_TEST_Battery}, 14
multipath-dense scenarios, 4 jamming scenarios of different power levels, and
3 scenarios exhibiting negligible interference beyond thermal and multi-access
noise.  The details of the data sets, including their source, are given Table II
of~\cite{wesson2018pincer}.  To ensure a fair and direct comparison with the PD
detector, these data were pre-processed using the same approach as
in~\cite{wesson2018pincer}.

\section{Experimental Results}
\label{sec-results}
\begin{figure}[t]
\centering
\includegraphics[width=0.5\textwidth]{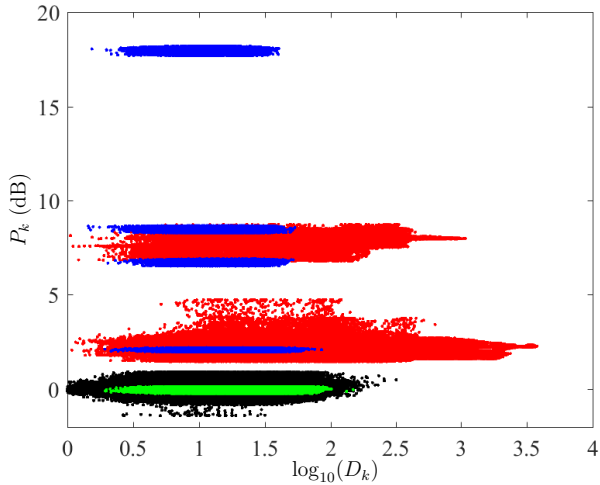}
\caption{Empirical distortion $D_k$ and received power $P_k$ measurements for
  interference-free (green), multipath (black), spoofing (red), and jamming
  (blue) from the 27 experimental recordings.}
\label{exp41taps}
 \end{figure}
 This section presents the experimental performance assessment of the PD-ML
 detector based on the decision regions shown in Fig. \ref{fig:regions} being
 applied to the observations, shown in Fig. \ref{exp41taps}, derived from the
 27 experimental data recordings.  Table~\ref{tab:misclass1} summarizes the
 PD-ML detector's overall performance against the experimental data in terms
 of classification statistics.

\begin{table}[h]
	\centering
	\caption{As Table \ref{tab:misclassSim1}, but for the detector applied
          to the experimental recordings (\textit{sans} {\tt tb7}).}
	\begin{tabular}[c]{ccccc}
		\toprule
		Decision & \multicolumn{4}{c}{True Scenario} \\
		\cmidrule(r){1-1} \cmidrule(r){2-5}
		& $H_0$ & $H_1$ & $H_2$ & $H_3$ \\  \midrule
		$H_0$ & 0.8500& 0.4337 & 0.0001 & 0 \\
		$H_1$ & 0.1500 & 0.5632 & 0 & 0 \\
		$H_2$ & 0 & 0.0031 & 0.8289 & 0.2176 \\
		$H_3$ & 0 & 0 & 0.1676 & 0.7824\\
		\bottomrule
	\end{tabular}
	\label{tab:misclass1}
\end{table}

When comparing this table to the corresponding one for the PD detector (Table
III in~\cite{wesson2018pincer}), the important result is that classification
performance accuracy under a spoofing attack is improved more than four-fold.
That is, with the PD detector, spoofing attacks were classified as jamming for
~82\% of experimental spoofing detection trials. This is reduced to only 17\%
with the PD-ML detector.  All but 0.01\% of spoofing or jamming attacks result
in an alarm.  Compared to the PD-ML detector, there is nearly a two-fold
decrease in spoofing false alarms under $H_1$ (multipath).  Finally, the PD-ML
detector correctly classifies $56\%$ of the multipath data, compared to only
12\% with the PD detector.

Thus, although the PD-ML detector does not enjoy any significant advantage
over the PD detector when tested against Monte-Carlo-simulated data (Table
\ref{tab:misclassSim1}), it significantly outperforms the PD detector when
tested against the empirical dataset.  A likely explanation is that the PD-ML
detector's greater number of correlation taps (for Table \ref{tab:misclass1},
$l = 11$) allow spoofer-induced distortion to be readily detected over a wider
range of spoofer-to-authentic code phase offset. 

The PD-ML detector was also applied to the especially stealthy { \tt tb7}
attack~\cite{texbat_ds7_ds8}, which mounts a nulling attack during the beginning of the attack
before pull-off. This attack would be extremely difficulty to realize outside
the laboratory.  Soon after pull-off begins, the PD-ML detector correctly
classifies the attack as spoofing. Against the attack portion of the scenario,
its decision rates were $H_0$:~17\%, $H_1$:~17\%, $H_2$:~66\%, and $H_3$:~0\%,
as compared to $H_0$: 14\%, $H_1$: 10\%, $H_2$: 70\%, and $H_3$: 6\% for the
PD detector.  The PD-ML detector's attack detection power is $10\%$ lower than
the PD detector's on this attack, but it never mis-attributes spoofing as
jamming. This can be attributed to especially subtle nature of this attack. In particular, a larger portion of the period prior to pull-off is labeled as multipath by the PD-ML detector due to the authentic signal being nulled.

\begin{figure*}[htb!]
\begin{subfigure}[h]{0.45\linewidth}
\includegraphics[ width=0.9\textwidth]{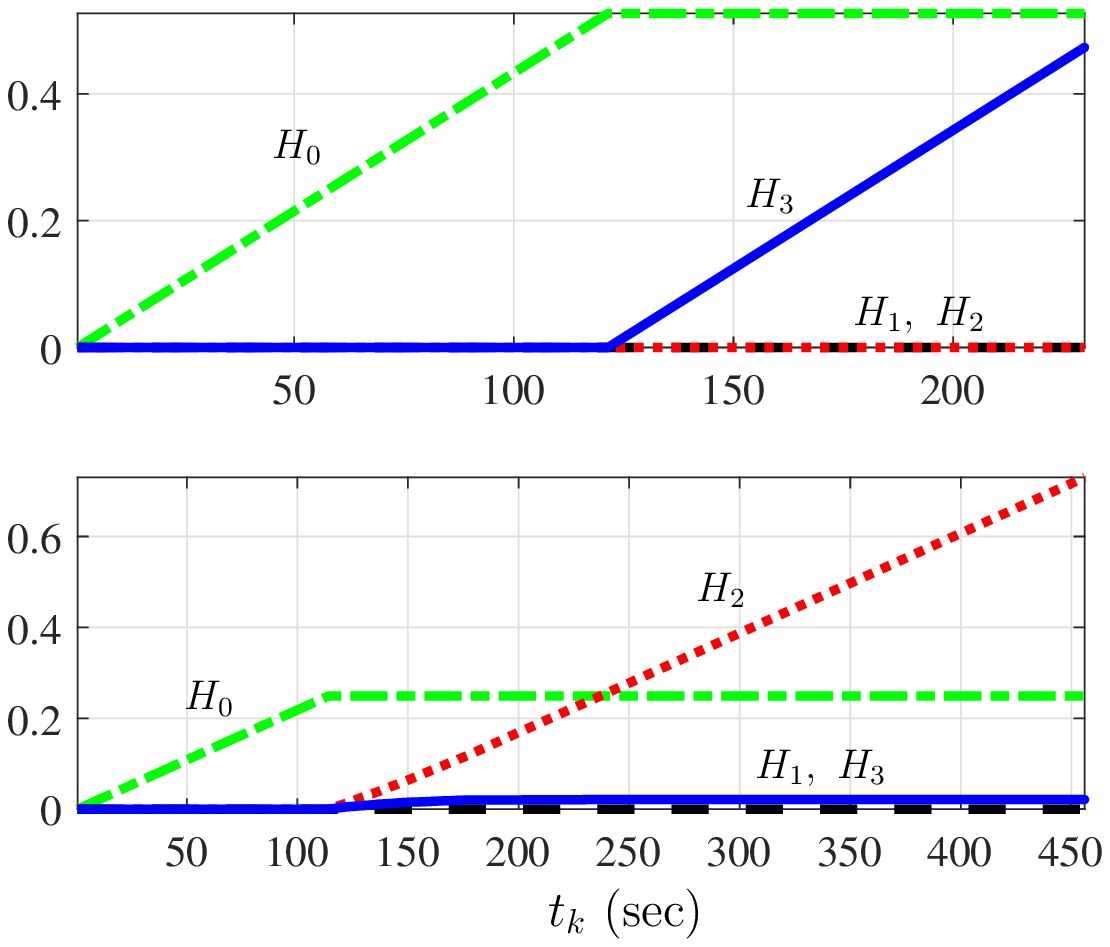}
\caption{Detection conducted with the proposed PD-ML detector.}
\label{detectionFig}
\end{subfigure}
\hfill
\begin{subfigure}[h]{0.5\linewidth}
 \includegraphics[height= 6 cm, width=0.8\textwidth]{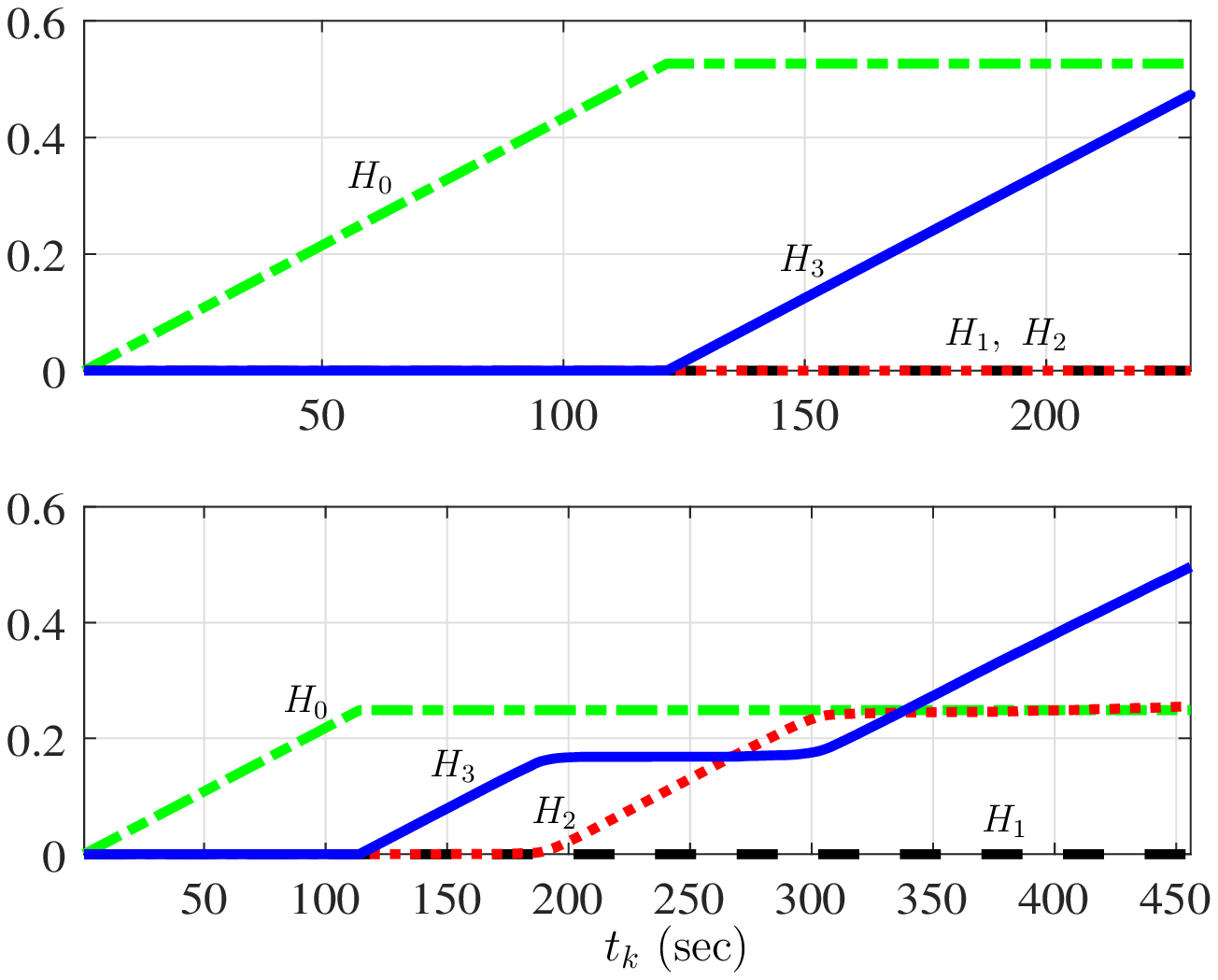}
  \caption{Detection conducted with PD detector. This figure has been reproduced from ~\cite{wesson2018pincer}  (Fig. 10) to allow for a side-by-side comparison.}
\label{detectionFigPD}
\end{subfigure}%
\caption{Cumulative time history of decisions for a single receiver tracking
  channel. Each trace represents the total number of times the corresponding
  hypothesis was chosen up to time $t_k$, normalized so that the final
  cumulative values sum to one. Both attacks begin at 120 seconds. Top:
  Jamming scenario {\tt jd3}. Bottom: Spoofing scenario {\tt tb4} with the
   detectors applied to a mid-elevation satellite signal.}
   \label{detectionFigOverall}

\end{figure*}

The PD-ML detector's superior spoofing vs. jamming classification is further
illustrated in of Fig. \ref{detectionFigOverall}, which shows the single-channel
cumulative time history of the detector's decisions for example jamming and
spoofing attack scenarios using both the proposed PD-ML detector and the original PD detector. Within Fig. \ref{detectionFigOverall}, \ref{detectionFigPD} shows the reproduced Fig. 10 from
~\cite{wesson2018pincer} and illustrates the performance of the PD detector. In Fig. \ref{detectionFig},  detection is assessed against the same attack scenarios using the proposed PD-ML detector.  Like the PD detector, in the jamming scenario (top
panels), the attack is detected immediately at its onset, and continuously
declared thereafter. In the spoofing scenario (bottom panels), the attack is
detected immediately, and briefly classified as jamming because the spoofer's
near-perfect initial code-phase alignment causes little distortion (and,
indeed, little harm to the receiver). However, as the spoofer begins its
pull-off, the PD-ML detector correctly declares the attack to be spoofing and
does so for the remainder of the attack. This is different from the PD
detector in Fig. \ref{detectionFigPD} , which declares the attack as an almost equal blend of jamming and
spoofing as the pull-off proceeds.

Finally, when comparing the PD-ML detector to the PD detector is it important to point out
that the PD-ML detector's distortion metric requires more computational complexity. In particular,
the PD detector's distortion metric simply subtracts two taps to determine symmetric difference~\cite{wesson2018pincer}. In contrast,
the PD-ML detector requires a Maximum-Likelihood estimation of the correlation function and requires access to
multiple tap data. Further, this estimator is executed multiple times. First over a coarse search for the code-phase and then
during a refinement stage via bisection.  As such, one should consider complexity against performance when deciding to
the chose  the PD-ML detector or the PD detector.

\section{Conclusions}
\label{sec:conclusions}
This paper presented the PD-ML detector, an extension to the recently-proposed
PD detector. The PD-ML detector employs a maximum-likelihood multipath
estimator and uses the magnitude of its post-fit residuals to monitor
distortion, as opposed to the PD detector's use of a symmetric
difference. Like the PD detector, the PD-ML detector traps a would-be attacker
between simultaneous monitoring of received power and complex correlation
function distortion.  In a head-to-head evaluation based on 27 high-quality
experimental recordings of attack and non-attack scenarios, the PD-ML detector
was shown to be significantly better at classifying spoofing vs. jamming,
exhibiting a nearly four-fold improvement.  In addition, the PD-ML detector
was shown to significantly improve the classification of multipath data. The
improved performance improvement comes at the expense of the additional
computational complexity associated with producing correlation products from a
larger number of taps and processing these with a maximum-likelihood
estimator.  Thus, depending on the application, the simpler PD detector may be
more favorable despite the better classification performance offered by the
PD-ML detector. In future work, it will be important to evaluate the the sensitivity of both the PD-ML and PD detectors on receiver's with a narrower front-end bandwidth (i.e., 2 MHz).

 The code for the PD-ML detector has been made available at the
same repository as the PD detector:
https://github.com/navSecurity/P-D-defense~\cite{pdDefenseRepo}.

\section*{Acknowledgements}
Jason Gross's work on this project was supported in part by a West Virginia
University Big XII Faculty Fellowship. Todd Humphreys's work on this project
has been supported by the National Science Foundation under Grant No. 1454474
(CAREER) and by the Data-supported Transportation Operations and Planning
Center (DSTOP), a Tier 1 USDOT University Transportation Center.

% ------------------------------------------------------------------------
\bibliographystyle{IEEEtran}
\bibliography{pangeaLocal}
% ------------------------------------------------------------------------
\end{document}